# Observation of Exciton Polariton Condensation in a Perovskite Lattice at Room Temperature

Rui Su[1,5], Sanjib Ghosh[1,5], Sheng Liu[1], Carole Diederichs[2,3], Timothy C.H. Liew[1,2]* & Qihua Xiong[1,2,4]*

[1]Division of Physics and Applied Physics, School of Physical and Mathematical Sciences, Nanyang Technological University, 637371, Singapore.

[2]MajuLab, International Joint Research Unit UMI 3654, CNRS, Université Côte d'Azur, Sorbonne Université, National University of Singapore, Nanyang Technological University, Singapore.

[3]Laboratoire de Physique de l'Ecole Normale Supérieure, ENS, Université PSL, CNRS, Sorbonne Université, Université de Paris, Paris, France.

[4]NOVITAS, Nanoelectronics Center of Excellence, School of Electrical and Electronic Engineering, Nanyang Technological University, 639798, Singapore.

[5]These authors contributed equally to this work.

*Corresponding author.  Email: Qihua@ntu.edu.sg (Q.X);  Timothyliew@ntu.edu.sg (T.L.)

**Abstract:**

Bose-Einstein condensation in strongly correlated lattices provides the possibility to coherently generate macroscopic quantum states, which have attracted tremendous attention as ideal platforms for quantum simulation to simulate complex many-body problems. Ultracold atoms in optical lattices[1,2] are one of such promising systems, where their realizations of different phases of matter exhibit promising applications in condensed matter physics, chemistry, and cosmology[3] . Nevertheless, this is only accessible with ultralow temperatures in the nano to micro Kelvin scale set by the typical inverse mass of an atom. Alternative systems such as lattices of trapped ions[4] and superconducting circuit arrays[5] also rely on ultracold temperatures. Exciton polaritons with extremely light effective mass[6], are regarded as promising alternatives to realize Bose-Einstein condensation in lattices at higher temperatures[7-12]. Along with the condensation, an efficient exciton polariton quantum simulator[13] would require a strong lattice with robust trapping at each lattice site as well as strong inter-site coupling to allow coherent quantum motion of polaritons within the lattice. A strong lattice can be characterised with a larger forbidden bandgap opening and a larger lattice bandwidth compared to the linewidth. However, exciton polaritons in a strong lattice have only been shown to condense at liquid helium temperatures[7-12]. Here, we report the observation of exciton polariton condensation in a one-dimensional strong lead halide perovskite lattice at room temperature. Modulated by deep periodic potentials, the strong lead halide perovskite lattice exhibits a large forbidden bandgap opening up to 13.3 meV and a lattice band up to 8.5 meV wide, which are at least 10 times larger than previous systems. Above a critical density, we observe exciton polariton condensation into $p_y$ orbital states with long-range spatial coherence at room temperature. Our result opens the route to the implementation of polariton condensates in quantum simulators at room temperature.

Microcavity exciton polaritons are part-light, part-matter bosons emerging from the quantum hybridization of excitons and microcavity photons. Inheriting strong nonlinearity from their excitonic fractions and low effective mass from their photonic parts, exciton polaritons provide the possibility to achieve Bose-Einstein condensation at much higher temperatures than those of typical ultracold atom cases. Above a critical density, exciton polaritons spontaneously occupy a low energy state to form polariton condensates with collective coherence. Along with the development of microfabrication techniques, one can precisely introduce periodic potentials to trap exciton polariton condensates, forming artificial atoms with sizeable scalability and controllability as solid-state analogues of ultracold atoms in optical lattices. In addition, by means of spin-orbit coupling[14], magnetic[12] and artificial gauge fields[15], such exciton polariton condensates in lattices could be manipulated into solitons[16] and topological edge modes[17], while other works showed the ordering of their spins[18] and topological charges[19]. With these advances, exciton polariton condensates in strong lattices are undoubtedly promising candidates for constructing quantum simulators towards room temperature operation[13,20,21].

Exciton polariton condensation was previously achieved within a one-dimensional weak periodical lattice potential induced by depositing periodic strips of a metallic thin film on the microcavity surface[8]. Hindered by the small exciton binding energy and shallow lattice potentials, one can only realize polariton condensation in lattices at liquid helium temperatures and the forbidden bandgap as the direct evidence of lattice potentials cannot be resolved. In recent years, there have been growing interests in utilizing wide bandgap semiconductors, such as GaN[22], ZnO[23] and organic semiconductors[24], as they can sustain stable exciton polariton condensation at room temperature[21]. Weak lasing in a weak lattice was demonstrated in a ZnO microcavity with smaller forbidden bandgap than the linewidth[25]. While polariton condensation in strong lattices at room temperature still remains challenging, recently, perovskite microcavities[26] have been shown to sustain stable exciton polariton condensation[27], significant polariton-polariton interactions[28], and long-range coherent polariton condensate flow[29] at room temperature. Along with the advantage of ease of fabrication[30], perovskite microcavities emerge as ideal candidates to overcome such limitations.

In our experiments, we create a one-dimensional micropillar array connected by channels in a cesium lead halide perovskite $\lambda$ microcavity characterized by a large vacuum Rabi splitting of ~120 meV and a negative detuning of ~176 meV. As shown in Fig.1a, the perovskite $\lambda$ microcavity consists of a 150-nm-thick perovskite layer between a 50-nm $SiO_2$ and a 61-nm poly (methyl methacrylate) (PMMA) spacers, sandwiched by 30.5 pair bottom $SiO_2/TiO_2$ distributed Bragg reflectors (DBR) and 12.5 pair $SiO_2/Ta_2O_5$ top DBR. Specifically, we create the micropillar array by merely patterning the spacer layer of PMMA with electron beam lithography, preserving the favorable polariton characteristics (Methods). Given the maturity of electron beam lithography technique, the use of PMMA as the spacer layer provides a huge degree of tunability in terms of pattern design and thickness control, which is crucial for a potential quantum simulator demanding sizeable scalability and controllability. As shown in Fig. 1b, the micropillar array consists of 10 pillars with diameter of 1.0 µm connected by 9 channels with a width of 0.5 µm. The center to center distance between the nearest pillars is 1.2 µm. Under the 61-nm thick PMMA layer, the induced deep periodic potential is as large as ~ 433 meV at normal incidence (See Supplementary Information for details), which suggests that polaritons are well-trapped in the lattices. When considering a single pillar, owing to the three dimensions of confinement in space, the photonic modes and thus the polariton modes will exhibit discrete energy levels where the ground state $s$ displays a cylindrical symmetry and the first excited state $p$ is two-fold degenerate with antisymmetric orbitals ($p_x$ and $p_y$) orthogonal to each other (Fig. 1c). Such confinement is well-confirmed by measuring the spatial image of the polariton emission along the diameter of a single pillar, showing the ground state *s* orbital at 2.232 eV and the first excited state *p_y* orbital at 2.273 eV as shown in Fig. 1d.

We study the polariton lattice behaviour first in the linear regime by momentum and real space imaging under a non-resonant continuous wave excitation at 2.71 eV (457 nm). Figure 2a depicts the momentum-space energy-resolved photoluminescence mapping imaged along the $k_y$ direction at $k_x = 0$, which agrees well with our theoretically calculated dispersion in Fig. 2g (Methods). It shows energy bands associated with fundamental modes of pillars and channels. The lower band originates from the hybridization between the *s* orbital states of the pillars and channel states, while the upper band mainly stems from the hybridization between the *p* orbital states of the pillars and channel states in the current energy scale region. It is worth mentioning that there is a missing band locates at E = 2.272 eV in Fig. 2a, compared with the theoretically calculated dispersion in Fig. 2g. We speculate that the population might be inhibited by selection rules in polariton-phonon or polariton-exciton scattering processes, as it corresponds to states with a different symmetry to the others. Detailed reasons need to be investigated. At the edge of first Brillouin zone at $k_y = 2.8$ $\mu m^{-1}$, we observe a complete band gap opening between the lower band and the upper band. Figure 2b displays the polariton photoluminescence spectrum with Gaussian function fittings taken at edge of the first Brillouin zone (red dashed line in Fig. 2a), from which we can clearly distinguish a dramatic energy splitting up to ~ 13.3 meV and a smaller linewidth of 6 meV. This large splitting value is at least 10 times larger than that of previous polariton lattice systems[10,11,17,31], which results from the small size of the micropillars (1.0 µm) and the induced deep potential (~ 433 meV) in our perovskite lattice system. Such large bandgap forbids inter-band transition due to any external perturbation and therefore provides a robust confinement of polaritons within the lattice sites. Along with the large bandgap, due to the introduction of the channels along the *y* axis, we also observe a large lattice bandwidth of ~ 8.5 meV, which indicates a large coupling strength (~

2.1 meV) between the nearest neighbour lattice sites. Such large inter-site coupling allows coherent motion of polaritons within the lattice sites.

To further elucidate the band structure of our lattice, we perform real space imaging at different energies (black horizontal dashed line in Fig. 2a) by using a narrow laser line filter. Figure 2c displays the real space image taken at E = 2.2312 eV in the lower band (black dashed line c in Fig. 2a), which clearly shows the non-degenerate cylindrical symmetry, confirming the *s* orbital mode nature. Figure 2d depicts the real space image taken at E = 2.2512 eV in the upper band (black dashed line d in Fig. 2a) where the emission is purely from the connecting channels, demonstrating the channel state nature. Figures 2e and 2f show the real space images taken at E = 2.2710 eV and E = 2.2870 eV in the upper band (black dashed line e and f in Fig. 2a), respectively. Both of them exhibit clear two-fold degenerate antisymmetric orbitals but orthogonal to each other, confirming their $p_x$ and $p_y$ orbital state nature. Ideally, in such a one-dimensional confined lattice based on an in-plane isotropic microcavity, polaritons would be more confined along the *x* axis than the *y* axis due to the opening of the channels along the *y* axis, which would result in higher energy of $p_x$ orbital states than that of $p_y$ orbital states. However, such scenario is in sharp contrast to our observation where $p_x$ orbital states locate at lower energies. The reversal of energy sequence between $p_x$ and $p_y$ orbital states is instead explained by the birefringent behaviour in an anisotropic perovskite structure[32], which results in anisotropic polaritons possessing smaller effective mass along *y* axis than polaritons along *x* axis. Such real-space behaviour can be well reproduced in theoretical calculations at similar energies (± 1 meV, blacked dashed lines in Fig. 2g) as shown in Fig. 2i to Fig. 2l (Methods).

In order to study our lattice system in the nonlinear regime, we use a 400 nm pulsed laser with a pulse duration of 100 fs and repetition rate of 1 kHz which excites the whole lattice. Under low excitation power of 0.7 $P_{th}$ in Fig. 3a, the polariton dispersion exhibits the band structure of our lattice system in accordance with Fig. 2a. Under strong excitation of 2.0 $P_{th}$, triggered by stimulated bosonic scattering, polaritons tend to condense at selected states with a maximum gain. As shown in Fig. 3b, we observe a macroscopic occupation of polaritons in the upper band characterized by a narrow momentum distribution and an intensity increase by three orders of magnitude, which suggests the occurrence of polariton condensation in our lattice system. The energy-resolved spatial image of the lattice in Fig. 3c clearly shows two-fold degenerated lobes in each pillar, suggesting the nature of polariton condensation into $p_y$ orbital states. To characterize the transition quantitatively, we plot the evolution of the emitted photon flux in the condensation state as a function of the pump fluence in Fig. 3d. A clear superlinear increase of the emitted photon flux by three orders of magnitude is observed when the pump fluence crosses the threshold $P_{th}$ ($P_{th}$ = 15 µJ/cm$^2$). Simultaneously, as shown in Fig. 3e, the emission linewidth correlatedly narrows by a factor of six at the threshold, suggesting the spontaneous build-up of the temporal coherence in the condensation regime. With the increase of pumping fluence as shown in Fig. 3f, the polariton emission exhibits a continuous blueshift behaviour resulted from repulsive nonlinear interactions, serving as another crucial evidence of polariton condensation in our lattice system.

The occurrence of exciton polariton Bose-Einstein condensation manifests as the phase transition from thermal phases to a quantum condensed phase, which is accompanied by spontaneous build-up of collective coherence. Particularly, the spontaneous appearance of off-diagonal long-range order is one of the defining features for exciton polariton condensate formation. This can be evidenced by demonstrating extended spatial coherence $g^{(1)}(r)$ across the condensates. In the thermal phase regime below the threshold, the polaritons are expected to possess short-range phase coherence with correlation length limited by the thermal de Broglie wavelength of ~1 µm. While in the quantum condensed phase regime above the threshold, polariton condensates will exhibit long-range phase coherence, covering the whole condensate region. To demonstrate the emergence of long-range spatial coherence, we send the real space image of polariton condensates into a Michelson interferometer with one arm replaced by a retroreflector which allows image inversion in a centrosymmetric way. The interference fringe contrast is the manifestation of phase coherence between points *r* and *-r* with respect to the center. As shown in Fig. 4a, when the real space image of polariton condensates at pump fluence of 2.0 $P_{th}$ (top panel) superimposes with its reverted image (mid panel), clear interference fringes can be well-resolved throughout the whole lattice region with distance separation as large as 12 µm, evidencing the build-up of the long-range spatial coherence in our lattice system. It is worth mentioning that the interference fringes are shifted by a half period in the middle of the interference pattern (Fig. 4b), suggesting a $\pi$ phase shift. This is caused by the minor displacement between two interfering images which leads to simultaneous overlapping of a lobe with the other two lobes with $\pi$ phase difference (See Supplementary Information for details).

In conclusion, we have realized a strongly coupled polariton lattice based on a perovskite microcavity at room temperature. The deep periodic potential up to hundreds of meV enables a giant band gap opening up to 13.3 meV, which is one order of magnitude larger than previous GaAs systems. By optical pumping, we are able to create an array of polariton condensates in the $p_y$ orbital state with long-range spatial coherence at room temperature. Our results reveal a brand-new polariton lattice system that can support polariton condensation at room temperature and simultaneously possess sizeable tunability in terms of potential landscape engineering and lattice design. We anticipate that our work will open the way to polariton lattice based topological polaritonic devices and quantum simulators operating at room temperature.

**Acknowledgements** Q.X. acknowledges the support from the Singapore National Research Foundation through the NRF Investigatorship Award (NRF-NRFI2015-03) and the Singapore Ministry of Education via AcRF Tier 2 grant (MOE2015-T2-1-047) and Tier 1 grants (RG103/15 and RG113/16). T.C.H.L. acknowledges the support of the Singapore Ministry of Education via AcRF Tier 2 grant (MOE2017-T2-1-001).

**Author contribution** R.S. fabricated the device and performed all the optical measurements. S.G. performed the theoretical calculations. S.L. conducted the atomic force microscopy measurements. R.S., S.G., T.C.H.L. and Q.H.X. analysed the data and wrote the manuscript, with input from all the authors.

**Competing interests** The authors declare no competing interests.

# METHODS

**Perovskite Microcavity Fabrication.** The bottom DBR is fabricated with an electron beam evaporator, consisting of 30.5 pairs of titanium oxide and silicon dioxide. A 50 nm-thick silicon dioxide is then deposited on the bottom DBR as the spacer layer. The 150 nm-thick perovskite layer is transferred onto the bottom DBR layer by a dry-transfer method using scotch tape. The growth of all-inorganic cesium lead halide perovskite is described in our previous reports[30]. In the following, a 58 nm-thick PMMA spacer layer is spin-coated on the perovskite and patterned by e-beam lithography. The substrate is finally put in the e-beam evaporator to complete the fabrication of the top DBR which consists of 12.5 pairs of silicon dioxide and tantalum pentoxide.

**Optical spectroscopy characterization.** The momentum-space and real-space photoluminescence imaging are conducted using a home-built microphotoluminescence setup with Fourier imaging configuration. The emission from the perovskite microcavity is collected through a 50 × objective (NA= 0.75) and sent to a 550-mm focal length spectrometer (HORIBA iHR550) with a grating of 600 lines/mm and a liquid nitrogen–cooled charge coupled device of 256×1024 pixels. In the linear region, the perovskite microcavity is pumped by a continuous-wave laser (457 nm) with a pump spot of ~ 10 µm and the real space image is measured with a narrow laserline filter (linewidth of 1 nm, Semrock) on the detection path. In the nonlinear regime, the perovskite microcavity is pumped by a 400 nm off-resonant pulsed excitation with a pump spot of ~15 µm. The pulse duration is 100 fs and the repetition rate is 1 kHz.

**Theoretical Calculations.** The photon field $\varphi(r,t)$ and the exciton field $\chi(r,t)$ are described by the coupled equations:

$$i\hbar \frac{\partial \varphi}{\partial t}(r,t) = \left[ -\frac{\hbar^2}{2m_y}\frac{\partial^2}{\partial y^2} - \frac{\hbar^2}{2m_x}\frac{\partial^2}{\partial x^2} + U(r) \right] \varphi(r,t) + \frac{g_0}{2}\chi(r,t) \quad (1)$$

$$i\hbar \frac{\partial \chi}{\partial t}(r,t) = U_{ex}\chi(r,t) + \frac{g_0}{2}\varphi(r,t) \quad (2)$$

where $m_x$ and $m_y$ are the effective photon masses along two axes ($x$ and $y$) of the planar microcavity. $U(r)$ is representing the potential profile experienced by the photons that defines the one-dimensional lattice that we consider. The photon-exciton coupling strength $g_0$ is taken as 120 meV for our calculation. For the exciton field, we considered that the effective mass is extremely large compared to the photon masses, such that the kinetic term in Eq. 2 can be neglected. $U_{ex}$ is a constant potential experiencing by the exciton field. Here we considered a one-dimensional lattice where each lattice point is a circular potential well of diameter 1.0 µm and the centre to centre distance between two nearby potential wells is 1.2 µm. The nearby potential wells are connected through a channel of width 0.5 µm. The potential U is defined as 2664.0 meV outside the lattice potential and 2231.0 meV inside the lattice potential and $U_{ex}$ = 2407.7 meV, $m_y/m_x$ = 0.625 where $m_x$ = 2.55×10$^{-5}$ $m_e$ and $m_y$=1.6×10$^{-5}$ $m_e$ (here $m_e$ is the electron mass). Recasting Eqs. 1 and 2 into a coupled time independent Schrodinger equation, we solve numerically to find the eigenvalues $E_n$ and the eigenmodes $[\varphi_n(r)\ \chi_n(r)]$ where $\varphi_n(r)$ and $\chi_n(r)$ are the photonic and excitonic parts of the eigenmodes.

$$\begin{pmatrix} H & g_0/2 \\ g_0/2 & V_{ex} \end{pmatrix} \begin{pmatrix} \varphi(r) \\ \chi(r) \end{pmatrix} = i\hbar \frac{\partial}{\partial t}\begin{pmatrix} \varphi(r) \\ \chi(r) \end{pmatrix}$$

For extracting the dispersion of the photonic part of the Hamiltonian, we introduce a photon intensity at energy $E$ and component $y$ of momentum $k_y$ as,

$$I(k_y, E) = \frac{1}{\pi\sigma\delta} \sum_n \sum_{p_x} \left| \tilde{\varphi}_n(p_y, p_x) \right|^2 \exp\left[ -\frac{(E-E_n)^2}{\sigma^2} - \frac{(k_y-p_y)^2}{\delta^2} \right]$$

where $\tilde{\varphi}_n(p_y, p_x)$ is the Fourier transform of $\varphi_n(r)$ and $\sigma$ and $\delta$ are the linewidths in energy and momentum respectively. We considered $\sigma = 2$ meV and $\delta = 0.3$ $\mu$m$^{-1}$ for our calculation. The real space image of a mode with energy $E_n$ is given by $|\varphi_n(r)|^2$. For demonstration, we show four modes in real space with $E_n$ = 2.2306 eV, 2.2516 eV, 2.2715 eV, 2.2859 eV, as shown in Fig. 2i to Fig. 2l.

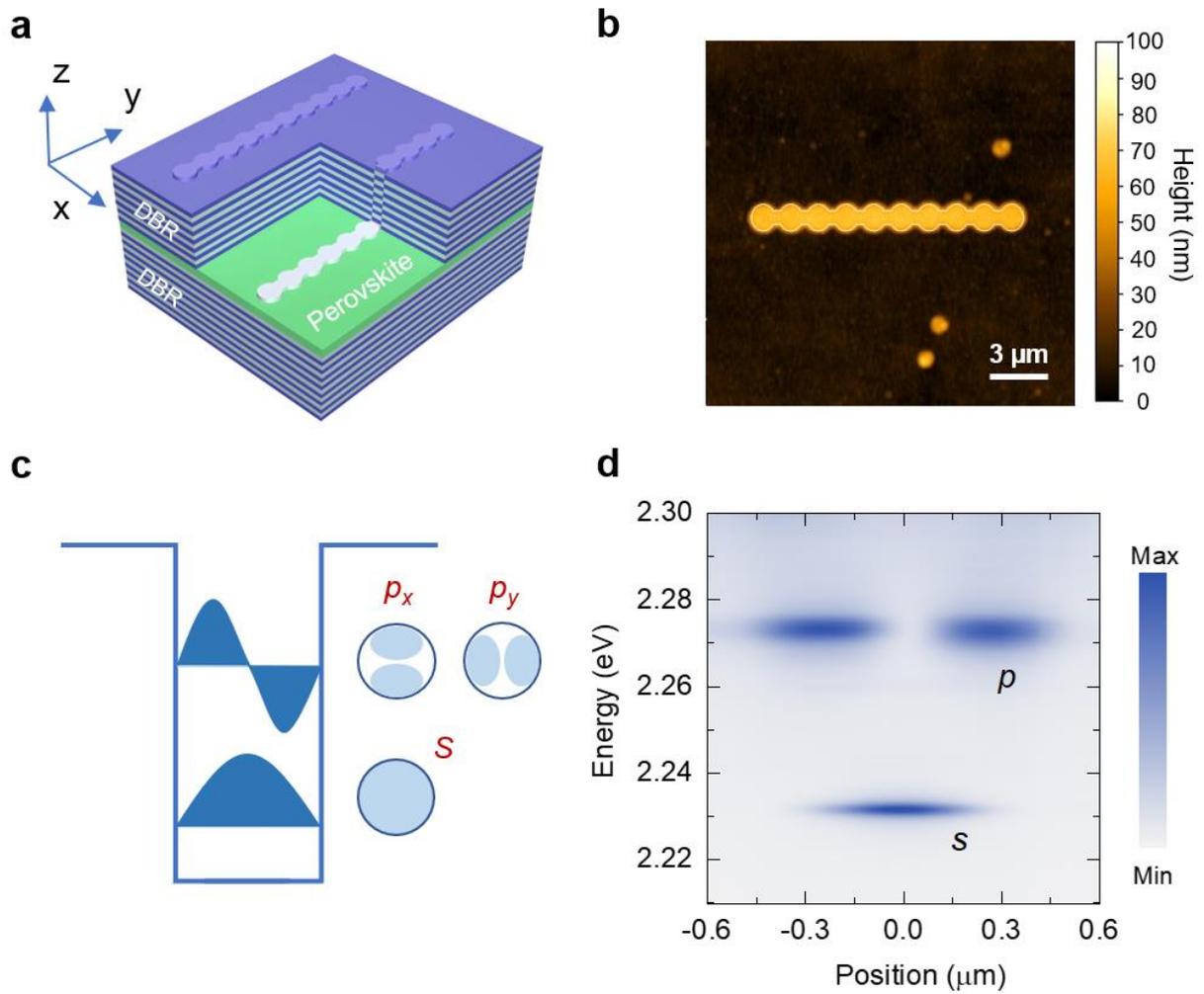

**Fig. 1 | Schematic diagram and characterization of the one-dimensional perovskite lattice.**
**a**, Structure schematic representation of the one-dimensional perovskite lattice, where the lattice pattern is created by patterning the PMMA spacer layer. **b**, Atomic force microscopy image of the one-dimensional perovskite lattice, showing 10 pillars with diameter of 1.0 µm connected by channels with width of 0.5 µm. The center to center distance of the nearest pillars is 1.2 µm. **c**, Schematic representation of orbital states in a single pillar, where the ground state *s* exhibits a cylindrical symmetry and the first excited state *p* is twice degenerate with antisymmetric orbitals ($p_x$ and $p_y$) orthogonal to each other. **d**, Energy-resolved spatial image of a single pillar along the diameter direction, showing a non-degenerate *s* state and a two-fold degenerate *p* state.

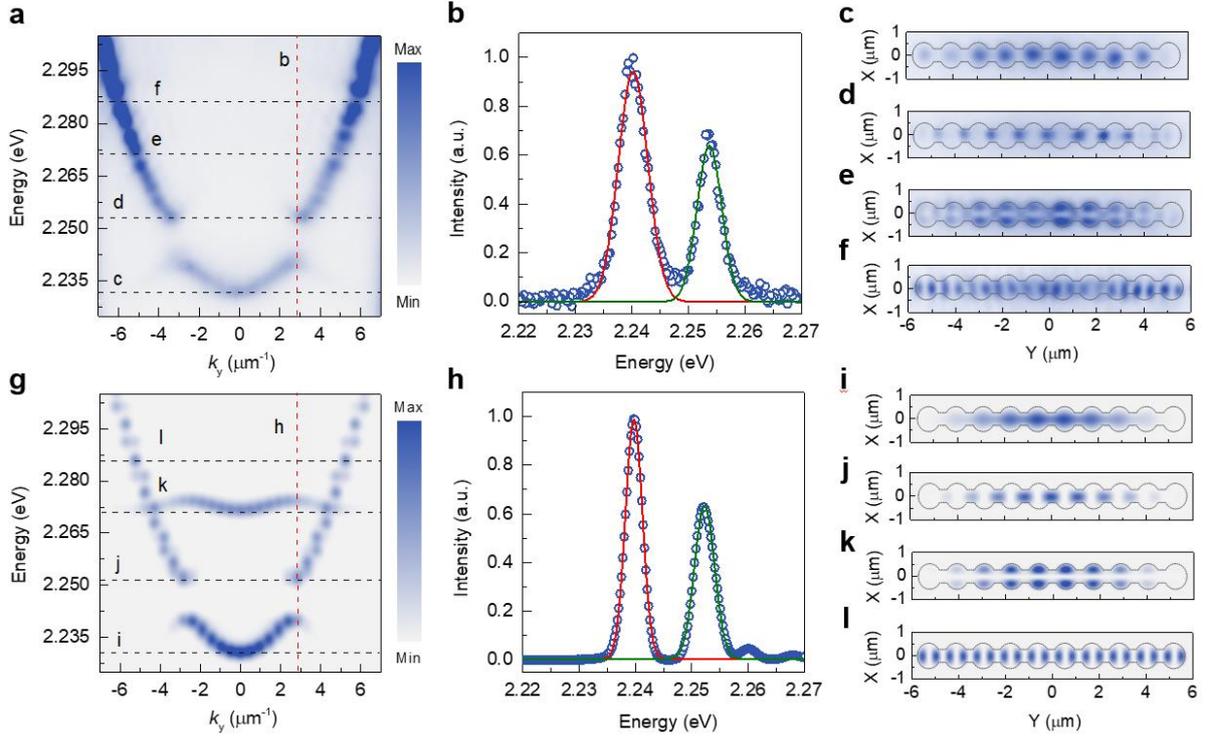

**Fig. 2 | Momentum-space and real-space imaging of the one-dimensional perovskite lattice at room temperature. a,** Experimental momentum-space polariton dispersion at $k_x = 0 \ \mu m^{-1}$ as a function of $k_y$, showing a large forbidden bandgap of 13.3 meV and a large bandwidth of 8.5 meV. The black dashed lines represent the energy selections for real-space imaging. The red dashed line represents the edge of the first Brillouin zone. **b,** Experimental polariton emission spectrum fitted with Gaussian function at the edge of the first Brillouin zone ($k_y = 2.8 \ \mu m^{-1}$), showing dramatic energy splitting as large as 13.3 meV between the upper and the lower bands. **c-f,** Experimental real-space images of the perovskite lattice at energies of 2.2312 eV, 2.2512 eV, 2.2710 eV, 2.2870 eV, corresponding to the black dashed lines in Figure **a**. **g,** Theoretical calculated momentum-space polariton dispersion at $k_x = 0 \ \mu m^{-1}$ as a function of $k_y$. **h,** Theoretical calculated polariton emission spectrum fitted with Gaussian function at the edge of the first Brillouin zone ($k_y = 2.8 \ \mu m^{-1}$), showing dramatic energy splitting as large as 11 meV between the upper and the lower bands. **i-l,** Theoretical calculated real-space images of the perovskite lattice at energies of 2.2306 eV, 2.2516 eV, 2.2715 eV, 2.2859 eV, corresponding to the black dashed lines in Figure **g**.

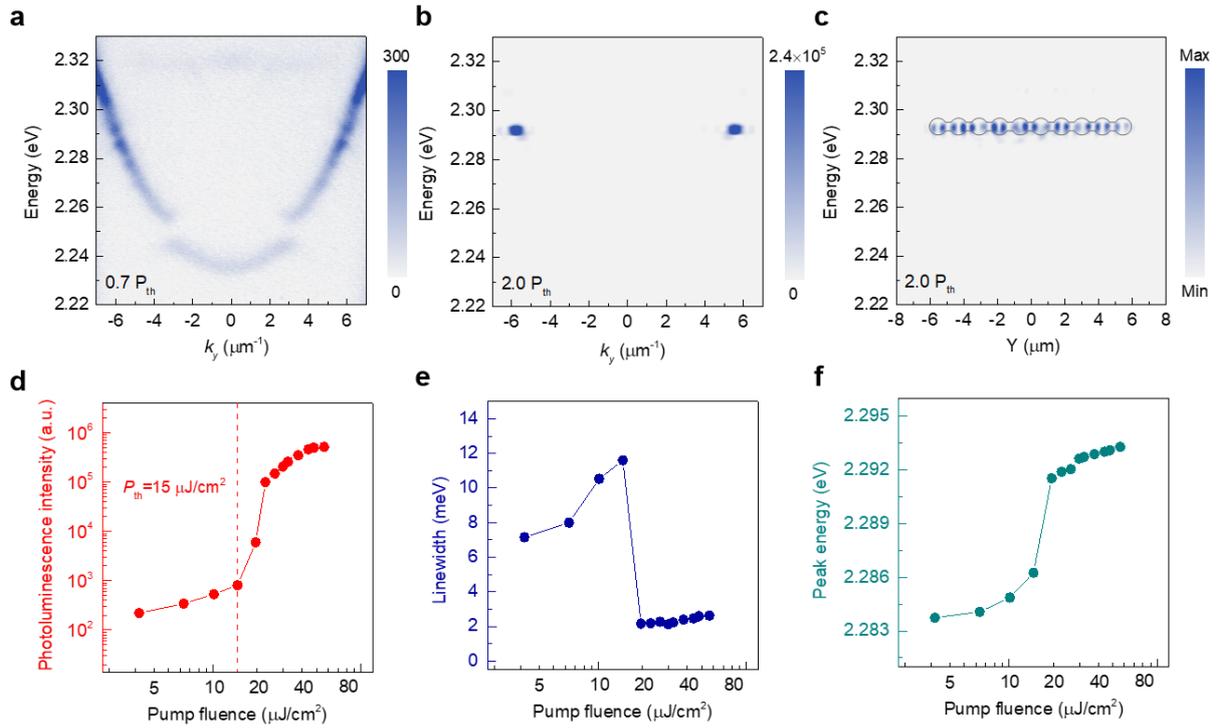

**Fig. 3 | Characterization of exciton polariton condensation in the one-dimensional perovskite lattice at room temperature. a**, Polariton dispersion in the thermal regime at the pump fluence of 0.7 $P_{th}$, revealing the band structure of the perovskite lattice. **b**, Polariton dispersion in the condensation regime at the pump fluence of 2.0 $P_{th}$, showing a macroscopic occupation of polaritons in the upper band with a narrow momentum distribution and an intensity increase by three orders of magnitude. **c**, Energy-resolved spatial image of the perovskite lattice in the condensation regime at the pump fluence of 2.0 $P_{th}$, showing clear twice-degenerate lobes in each pillar, which suggests the nature of polariton condensation into $p_y$ orbital states. **d**, Emitted photon flux in the condensation state as a function of pump fluence in a log-log scale, demonstrating a clear superlinear increase trend by three orders of amplitude. **e**, Evolution of the emission linewidth in the condensation state as a function of pump fluence, exhibiting dramatic collapse from 12 meV to 4 meV at the threshold. **f**, Evolution of the emission peak energy in the condensation state as a function of pump fluence, displaying a continuous blueshift trend.

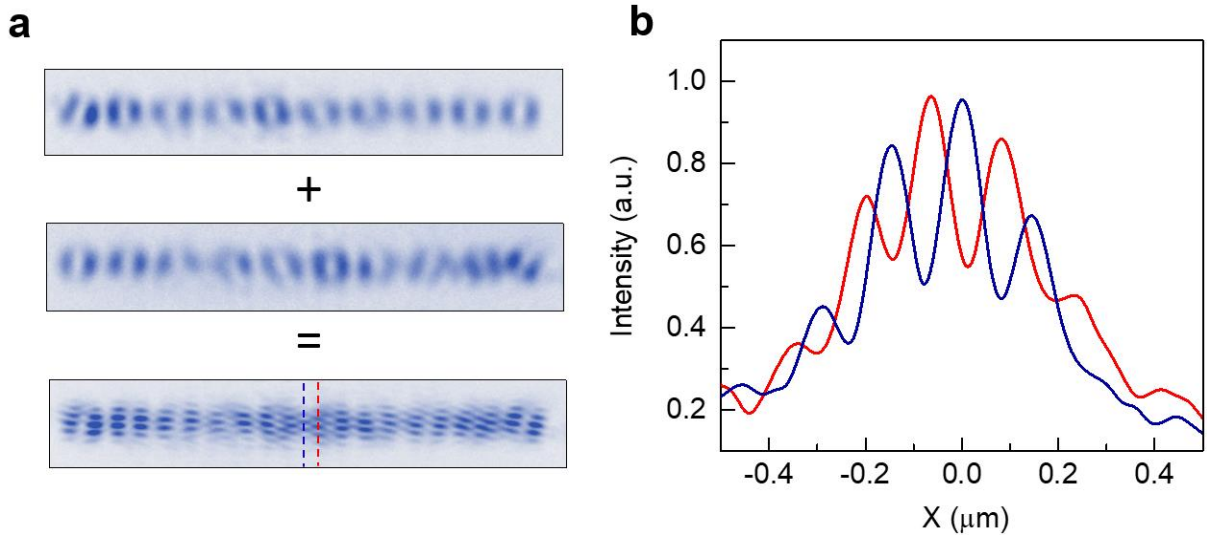

**Fig. 4 | Build-up of long-range spatial coherence in the condensation regime of the one-dimensional perovskite lattice at room temperature. a**, Superposition of the real space image and its inverted image of the perovskite lattice at 2.0 $P_{th}$, where clear interference fringes are readily identified within a distance as large as 12 μm, demonstrating the build-up of the long-range spatial coherence. The interference fringes are shifted by a half period in the middle of the interference pattern, which is caused by simultaneous overlapping of a lobe with the other two lobes with $\pi$ phase difference. **b**, Interference spectra corresponding to red and blue dashed lines taken from the interference pattern in Figure **a**, confirming the $\pi$ phase difference of adjacent lobes.